%% file: aajets_main.tex
\documentclass[12pt]{article}

\pdfoutput=1
\usepackage{jheppub}
\usepackage{graphicx,epsfig,amsmath,amssymb,booktabs}
\usepackage{caption,subcaption}
\usepackage{xspace}

\newcommand{\aapa}{\ensuremath{\gamma\gamma}}
\newcommand{\shortequal}{\ensuremath{\!\!\!=\!\!\!}}
\newcommand{\pT}{\ensuremath{p_\text{T}}}
\newcommand{\order}{\ensuremath{\mathcal{O}}}

\newcommand{\Sherpa}{S\scalebox{0.8}{HERPA}\xspace}
\newcommand{\GoSam}{G\scalebox{0.8}{O}S\scalebox{0.8}{AM}\xspace}

\newcommand{\QGraf}{QG\scalebox{0.8}{RAF}\xspace}
\newcommand{\FORM}{F\scalebox{0.8}{ORM}\xspace}
\newcommand{\Spinney}{S\scalebox{0.8}{PINNEY}\xspace}
\newcommand{\Ninja}{N\scalebox{0.8}{INJA}\xspace}
\newcommand{\Samurai}{S\scalebox{0.8}{AMURAI}\xspace}
\newcommand{\GolemNF}{G\scalebox{0.8}{OLEM}95\xspace}
\newcommand{\OneLoop}{O\scalebox{0.8}{NE}L\scalebox{0.8}{OOP}\xspace}
\newcommand{\FastJet}{F\scalebox{0.8}{AST}J\scalebox{0.8}{ET}\xspace}

\newcommand{\DIPHOX}{D\scalebox{0.8}{IPHOX}\xspace}
\newcommand{\MCFM}{M\scalebox{0.8}{CFM}\xspace}
\newcommand{\twogammaMC}{2\scalebox{0.8}{gammaMC}\xspace}
\newcommand{\VBFatNLO}{V\scalebox{0.8}{BF@NLO}\xspace}

\title{Electroweak corrections to diphoton plus jets}

\author[a]{Mauro Chiesa,}
\author[b]{Nicolas Greiner,}
\author[b]{Marek Sch\"onherr,}
\author[c]{Francesco Tramontano}

\affiliation[a]{Institut f\"ur Theoretische Physik und Astrophysik, Julius-Maximilians-Universit\"at W\"urzburg,
  Emil-Hilb-Weg 22, D-97074 W\"urzburg, Germany} 
\affiliation[b]{Physik Institut, Universit{\"a}t Z{\"u}rich, Winterthurerstr.190, 8057 Z\"urich, Switzerland}
\affiliation[c]{Dipartimento di Fisica, Universit\`a di Napoli ``Federico II'' and INFN,
  sezione di Napoli, via Cintia, 80126 Napoli, Italy}

\preprint{
  \small
  \begin{flushright}
    ZU--TH 20/17 \\ MCNET--17--11
  \end{flushright}
}

\abstract{
  We calculate the next-to-leading order electroweak corrections to 
  the production of a photon pair in association with zero, one and 
  two jets at the LHC.
  We use \GoSam and \Sherpa to obtain the results in a fully 
  automated way. For a typical set of fiducial cuts the electroweak 
  corrections lead to a modification 
  of the total cross section of up to $3\%$, depending 
  on the jet multiplicity.  We find substantial contributions in differential distributions, leading to tens of per cent corrections
  for phase space regions within the reach of the LHC. Furthermore we investigate the importance of photon induced processes
  as well as subleading contributions. Photon induced processes are found to be negligible, subleading contributions can 
  have a sizeable impact however they can be removed by appropriate phase space cuts.
}

\keywords{EW corrections, Photon, NLO, Jets}

\begin{document}

\maketitle

\section{Introduction}
\label{sec:intro}
The production of a pair of photons is an interesting and important process at the LHC. It played an essential role in the 
discovery of the Higgs boson~\cite{Aad:2012tfa,Chatrchyan:2012ufa} and in further refined measurements of fiducial cross sections
and differential observables~\cite{Aad:2014eha,Aad:2014lwa,CMS:2017nyv,CMS:2016ixj,Khachatryan:2015rxa}. Due to its clean experimental signal, 
diphoton production in hadron collisions offers a testing ground for perturbative quantum chromodynamics and has been widely
studied at the LHC~\cite{Aad:2011mh,Chatrchyan:2011qt,Aad:2012tba,Chatrchyan:2014fsa,Aaboud:2017vol}.
This process is also an important search channel for new physics (see~\cite{ATLAS:2016eeo,Aaboud:2016tru,Aad:2014gfa,Aad:2014ioa,
Khachatryan:2016yec,CMS:2016owr,CMS:2015dxe,Khachatryan:2015qba} for recent experimental searches). Diphoton signals
can occur in models with extra spatial components or in cascade decays of heavy new particles. Diphoton signals 
in association with jets and missing energy occur in gauge mediated SUSY scenarios.

Diphoton searches at the LHC are typically carried out fully inclusively in the number of jets. In fact, due to the 
hadronic environment there is a large probability that the photon pair will be accompanied by one or more high 
energetic jets. The two jet process is particularly interesting since it allows to asses the VBF production channel.

The next-to-leading (NLO) QCD corrections to diphoton production at hadron colliders have been computed in
Refs.~\cite{Aurenche:1985yk,Bailey:1992br} and they have been implemented in the fixed order
codes \DIPHOX~\cite{Binoth:1999qq}, \twogammaMC~\cite{Bern:2002jx}, \MCFM~\cite{Campbell:2011bn},
\VBFatNLO~\cite{Arnold:2008rz}. Theoretical predictions for $pp\to\gamma\gamma$ at QCD NLO matched with parton shower can be found
in Ref.~\cite{DErrico:2011cgc}, while in Ref.~\cite{Hoeche:2009xc} diphoton production is studied in the context of matrix element-parton shower merging.
Soft gluon resummation for $pp\to\gamma\gamma$ has been performed in~\cite{Balazs:2006cc,Balazs:1999yf,Cieri:2015rqa}. 
The diphoton process is known at NNLO in QCD~\cite{Catani:2011qz,Campbell:2016yrh}, at NLO in the electroweak 
coupling constant~\cite{Bierweiler:2013dja}. 
The diphoton plus one and two jet processes are known at next to leading order in
QCD~\cite{DelDuca:2003uz,Gehrmann:2013aga,Gehrmann:2013bga,Badger:2013ava,Bern:2014vza}.

In this paper we calculate the next-to-leading order electroweak 
corrections to the production of a pair of photons in association with up
to two jets. More precisely we calculate the electroweak corrections to the leading contribution that are of the order 
${\cal O}(\alpha^2\alpha_s^i)$ with $i = 0,1,2$, where $i$ denotes the number of jets. This calculation provides important results that are useful for measurements of a diphoton system in association with jets. They are particularly useful for background studies for Higgs production as well as in the context of boosted searches where one expects sizeable effects from electroweak corrections. In addition we asses the impact
of photon induced processes as well the impact of subleading electroweak contributions. The paper is organized as follows. In section
\ref{sec:setup} we describe the calculational setup before we discuss the results and the phenomenology of these processes
in section \ref{sec:results}. Finally we conclude in section \ref{sec:conclusions}.

\section{Calculational setup}
\label{sec:setup}
The calculation has been performed by combining the functionalities of the
\GoSam~\cite{Cullen:2011ac,Cullen:2014yla} and \Sherpa~\cite{Gleisberg:2008ta}
programs.
The \GoSam package generates the code for the numerical evaluation of
the one loop scattering amplitudes starting from the Feynman diagrams, 
generated with \QGraf~\cite{Nogueira:1991ex}, and writing them as 
$d$-dimensional integrand over the loop momentum, making use of 
\FORM~\cite{Vermaseren:2000nd,Kuipers:2012rf} and 
\Spinney~\cite{Cullen:2010jv} to perform the needed algebraic 
manipulations.
For the integrand reduction of the diagrams we use the \Ninja
library~\cite{Peraro:2014cba}, an implementation of the technique of integrand
reduction via Laurent expansion~\cite{Mastrolia:2012bu,vanDeurzen:2013saa}.
Alternatively one can choose other reduction strategies such as OPP reduction
method~\cite{Ossola:2006us,Mastrolia:2008jb,Ossola:2008xq} which is
implemented in $d$ dimensions in \Samurai~\cite{Mastrolia:2010nb}, or methods based on
tensor integral reduction as implemented in
\GolemNF~\cite{Heinrich:2010ax,Binoth:2008uq,Cullen:2011kv,Guillet:2013msa}.
We have used \OneLoop~\cite{vanHameren:2010cp} to evaluate the scalar integrals.

One loop electroweak renormalization is performed in the on-shell scheme
(see~\cite{Denner:1991kt} and references therein) within the 
complex-mass-scheme framework~\cite{Denner:1999gp,Denner:2005fg,Denner:2006ic}. 
Since for the processes under consideration the powers of $\alpha$
only come from  $\gamma f \overline{f}$ vertices, we use the $\alpha(0)$ 
input parameter scheme. 
All fermion masses, except for the top, are set to zero. 
This choice of input parameters leads to mass singularities both in the 
electric charge and in the photon wave function renormalization counterterms 
($\delta Z_e$ and $\delta Z_{AA}$), respectively, however these counterterms 
only appear in the combination $\delta Z_e + \frac{1}{2} \delta Z_{AA}$ 
where the mass singularities cancel.
Counterterm functions are first computed in conventional dimension 
regularization (CDR) and then converted in dimensional reduction (DRED) 
as detailed in \cite{Chiesa:2015mya}, to be used within the \GoSam framework.

The present paper is the first application of a new module in \GoSam 
for the fully automated generation of the electroweak renormalization 
counter terms. 

\Sherpa, on the other hand, is used for providing all tree-level matrix 
elements, infrared subtraction, process management and phase-space 
integration of all the contributions to the cross sections considered 
here \cite{Krauss:2001iv}. 
The infrared subtraction is carried out in the QED generalisation of 
the Catani-Seymour scheme~\cite{Catani:1996vz,Dittmaier:1999mb,Gleisberg:2007md,
  Kallweit:2014xda,Kallweit:2015dum,Kallweit:2017khh,Schonherr:2017xxx}
and includes the initial state collinear factorisation counter terms.
The two programs are interfaced through a dedicated interface based on the 
Binoth Les Houches Accord~\cite{Binoth:2010xt,Alioli:2013nda}. 
We have cross-checked the tree-level matrix elements and the renormalized 
pole terms of \GoSam against the tree-level matrix elements and the 
infrared pole terms of \Sherpa for several phase space points spanning 
multiple kinematic regimes and found excellent agreement.

\input{results}


\section{Conclusions}
\label{sec:conclusions}
Direct diphoton production channels constitute an important class
of background processes for
Higgs production in the Standard Model, where the Higgs subsequently decays
into a pair of photons, and for many other test of the SM and BSM searches. In this paper we calculated the
 next-to-leading order electroweak corrections
to the production of a photon pair in association with up to two jets. The calculation has been carried out fully automatically 
using the combination of Sherpa plus GoSam.  We restricted ourselves to the leading contributions, i.e the contributions of the 
orders of $\order(\alpha^2\alpha_s^i)$ with $i = 0,1,2$. For the total cross sections we found corrections between
$\sim+1\%$ and $-3\%$, depending on the jet multiplicity.  As expected the actual impact of the electroweak corrections 
become mostly effective in the high energy tail of the differential distributions like the transverse momentum, where one easily
obtains corrections of the order of $10-20\%$ for transverse momenta of the order of a few hundred GeV.  The inclusion of
next-to-leading order electroweak corrections is, therefore, important for a precise prediction of this class of processes.

We also assessed the importance of photon induced processes and find them to be negligible
which justifies the usage of pure QCD parton distribution functions.
For the $2$-jet case we also investigated the contributions of subleading and sub-subleading processes (i.e. contributions of
$\order(\alpha^3 \alpha_s)$ and $\order(\alpha^4 )$ respectively) at leading order accuracy. We found that the 
sub-subleading contributions supersede
the subleading contributions and lead to corrections that are of similar size as the electroweak corrections to the leading 
contribution. However, one should note that they are sizeable due to intermediate resonances that can be effectively cut away
by applying VBF cuts, in particular by a cut on the invariant mass of the dijet system.  We expect that they would lead to a 
negligible contribution in a VBF type analysis and the  ($\order(\alpha^2 \alpha_s^2)$) terms would yield
the dominant contribution.

\section*{Acknowledgements}
We would like to thank Jonas Lindert and Stefano Pozzorini for comparing results.
N.G.\ was supported by the Swiss National Science Foundation under contract
PZ00P2\_154829. M.S.\ was supported by PITN--GA--2012--315877 ({\it MCnet}) 
and the ERC Advanced Grant MC@NNLO (340983).

\providecommand{\href}[2]{#2}\begingroup\raggedright\endgroup

\end{document}

%% file: results.tex
\section{Numerical results}
\label{sec:results}

In this section we present numerical results for the next-to-leading 
order electroweak corrections in the production of (at least) two 
isolated photons, both inclusively and in association with at least one 
or two jets, $\text{pp}\to \gamma\gamma+0,1,2\,\text{jets}$ at the 
LHC at a centre-of-mass energy of 13\,TeV. 
Our calculation is performed in the Standard Model using the complex-mass 
scheme with the following input parameters
\begin{center}
  \begin{tabular}{rclrcl}
    $\alpha(0)$ &\shortequal& $1/137.03599976$  \qquad &&& \\
    $m_W$ &\shortequal& $80.385\; \text{GeV}$       & $\Gamma_W$ &\shortequal& $2.085\; \text{GeV}$ \\
    $m_Z$ &\shortequal& $91.1876\; \text{GeV}$      & $\Gamma_Z$ &\shortequal& $2.4952\; \text{GeV}$ \\
    $m_t$ &\shortequal& $171.2\; \text{GeV}$        & $\Gamma_t$ &\shortequal& $0$\;.
  \end{tabular}
\end{center}
The width of the top quark can be safely neglected as there are no 
diagrams containing it as an $s$-channel resonance which can potentially 
go on-shell. 
At the same time there are no diagrams 
containing the Higgs boson due to the absence of $W$, $Z$ or top quark 
propagators at LO.
All other parton masses and widths are set to zero, i.e.\ we are working 
in the five-flavour scheme.

Isolated photons are defined with the help of the smooth cone isolation 
criterion \cite{Frixione:1998jh} which limits the maximally allowed 
hadronic activity in a cone of size $R_{\gamma}$ around a photon to
\begin{equation}
  E_{{\rm had, max}} (r_{\gamma}) = \epsilon\, p_{T}^{\gamma} \left( \frac{1-\cos r_\gamma}
  {1-\cos R_\gamma}\right)^{n}\;,
  \label{eq:frix}
\end{equation}
where $r_{\gamma}$ denotes the angular separation between the photon and 
the parton. 
The free parameters $R_{\gamma}$, $\epsilon$, and $n$ are set to
\begin{equation}
R_{\gamma}=0.4\;, \quad \epsilon = 0.05\;, \quad n = 1\;. 
\end{equation}
Of the thus found photon candidates we require at least two to lie in the 
fiducial volume given by
\begin{equation}
 \label{eq:photon_cuts}
 p_{T_{\gamma_1}} > 40\; \text{GeV},\quad p_{T_{\gamma_2}} > 30\; \text{GeV}, \quad
 \left|\eta_{\gamma}\right| < 2.37\;.
\end{equation}
It is worth noting that at NLO EW it is possible to find more 
than two isolated photons, in which case any pair is allowed to 
fulfill these selection criteria. 
Of those photons, the one with largest transverse momentum is 
refered to as $\gamma_1$ in the following, while the one with 
second largest transverse momentum is refered to as $\gamma_2$. 
Any additional idenfied photons are kept as such. 
All identified photons are further required to be pairwise separated by 
$\Delta R(\gamma_i,\gamma_j)>0.4$, otherwise the event is discarded.

On the other hand, photons that are either not isolated according to 
eq.\ \eqref{eq:frix} or do not fulfill the cuts of 
eq.\ \eqref{eq:photon_cuts} are passed to the jet algorithm, 
along with all quarks and gluons of the event. 
Jet candidates are then found using the anti-$k_t$ clustering 
algorithm~\cite{Cacciari:2008gp} with a cone size of $R=0.4$ 
provided by the \FastJet package \cite{Cacciari:2011ma,Cacciari:2005hq}. 
The candidates are further required to have 
\begin{equation}
 p_{\text{T}_j} > 30\,\text{GeV},\quad \left|y_{j}\right| < 4.4\;.
\end{equation}

At this point we want to stress that the best procedure would be a 
democratic clustering \cite{Glover:1993xc}, where photons and partons 
are treated on the same footing and clustered by an infrared safe jet 
algorithm. 
However this necessitates the inclusion of additional components 
in the NLO calculation.
Indeed, one can have configurations where a hard photon is accompanied by a 
soft gluon forming a jet.
Such a configuration would lead to a QCD singularity which requires 
the virtual QCD corrections to 
a different underlying Born, where the gluon is replaced by a photon, 
to cancel this singularity. 
Thus if a photon and a gluon are clustered to a jet the gluon is 
required to carry a non-vanishing fraction of the momentum of the 
jet to avoid the occurrence of QCD infrared singularities. 
The difference to the scheme used in this paper is estimated to 
be very small \cite{Kallweit:2014xda}.
To ensure infrared safety wrt.\ possible soft gluons a photon
must not  carry a transverse 
energy fraction of more than $z_\text{thr}=0.5$ within a jet, 
if the jet comprises a photon and a gluon. 
This constitutes a less invasive version of the regularisation 
applied in \cite{Kallweit:2014xda,Kallweit:2015dum}.
A further requirement of $\Delta R(\gamma_i,j_j)>0.4$ removes 
the overlap between identified photons and jets.

The renormalization and factorization scales are chosen to be equal 
and set to
\begin{equation}
 \mu_{R,F} = \frac{1}{2}\sqrt{m_{\gamma \gamma}^2 + \hat{H}_\text{T}^2} 
 \qquad\text{with}\quad
 \hat{H}_\text{T} = \sum_i p_{\text{T},i}\;,
\end{equation}
wherein the sum runs over all partons but the identified photons 
\cite{Gehrmann:2013bga}. As our main focus in this paper is on the 
size of the corrections, i.e. the ratio with respect to the leading 
order and as we do not expect significant deviations from a variation,
we do not vary the scale but leave it fixed at the central value.

The invariant diphoton mass is formed using the two leading identified 
photons defined above.
We use the CT14nlo PDF \cite{Dulat:2015mca} set, interfaced through 
\textsc{Lhapdf6} \cite{Buckley:2014ana}, with its 
accompanying $\alpha_s$-parametrisation with $\alpha_s(m_Z)=0.118$ throughout. 
If not stated otherwise we neglect all photon initiated processes. Due to the smallness of 
the photon PDF and the fact that they enter only at NLO EW, their impact 
is estimated to be negligible. We will quantify this statement later in this section.


\begin{table}[h!]
  \centering
  \begin{tabular}{l|c|c|c}
      & $\;\;pp \to \gamma \gamma\;\;$
      & $\;\;pp \to \gamma \gamma +j\;\;$ 
      & $\;\;pp \to \gamma \gamma +j j\;\;$ \\
    \hline\hline
    $\sigma_\text{LO}\;\;[\text{pb}] $ & $5.398$ & $9.597$ & $2.073$\\
    \hline
    $\sigma_\text{NLO EW}\;\;[\text{pb}] $ & $5.449$ & $9.587$ & $2.009$\\
    \hline
    $\delta_\text{EW}\;\;[\%]$ & $\textcolor{white}{-}0.9$ & $-0.1$ & $-3.1$\\
  \end{tabular}
  \caption{
    Total cross sections at LO and NLO for $\aapa+0,1,2$\,jets 
    production at 13\,TeV at the LHC.
    \label{tab:xsec}
  } 
\end{table}

The total cross sections for the three different processes are listed 
in Tab.\ \ref{tab:xsec}. 
The first interesting point is the fact that, at 
leading order and next-to-leading order electroweak, the fiducial cross 
section for the $1$-jet process 
is almost a factor of two bigger than the $0$-jet process.
This is due to the quark-gluon channel opening up for the $1$-jet case 
whereas the $0$-jet process is mediated only by $q\bar{q}$ initial states. 
For the $2$-jet process there is then also the gluon-gluon initial state 
possible. 
This leads to the fact that the $2$-jet process is only roughly a factor 
of two smaller than the $0$-jet process although it is suppressed by two 
additional powers of $\alpha_s$. 
This is remedied by the inclusion of higher order QCD correction 
\cite{Catani:2011qz}, but is beyond the scope of this paper.
The NLO EW corrections to the total cross section are tiny, for the 
$0$-jet process they are positive and lead to an enhancement of one per cent. 
Increasing the jet multiplicity reduces the EW corrections and they 
become negative, for the $1$-jet process the cross section is essentially 
unchanged, for the $2$-jet process we find a reduction of three per cent.

We now turn to the discussion of a selection of the differential distributions where 
one expects to see a larger impact of the EW corrections than for the 
total cross sections, predominantly at large $\pT$. 
For the $2$-jet process we additionally include two different subleading 
contributions at tree-level: 'NLO EW$^*$' includes also 
$\order(\alpha_s\alpha^3)$ contributions, and 'NLO EW$^{**}$' includes both the 
$\order(\alpha_s\alpha^3)$ and $\order(\alpha^4)$ terms.

\begin{figure}[t!]
  \begin{center}
   \small
    \begin{tabular}{ccccccc}
      \includegraphics[width=.20\textwidth]{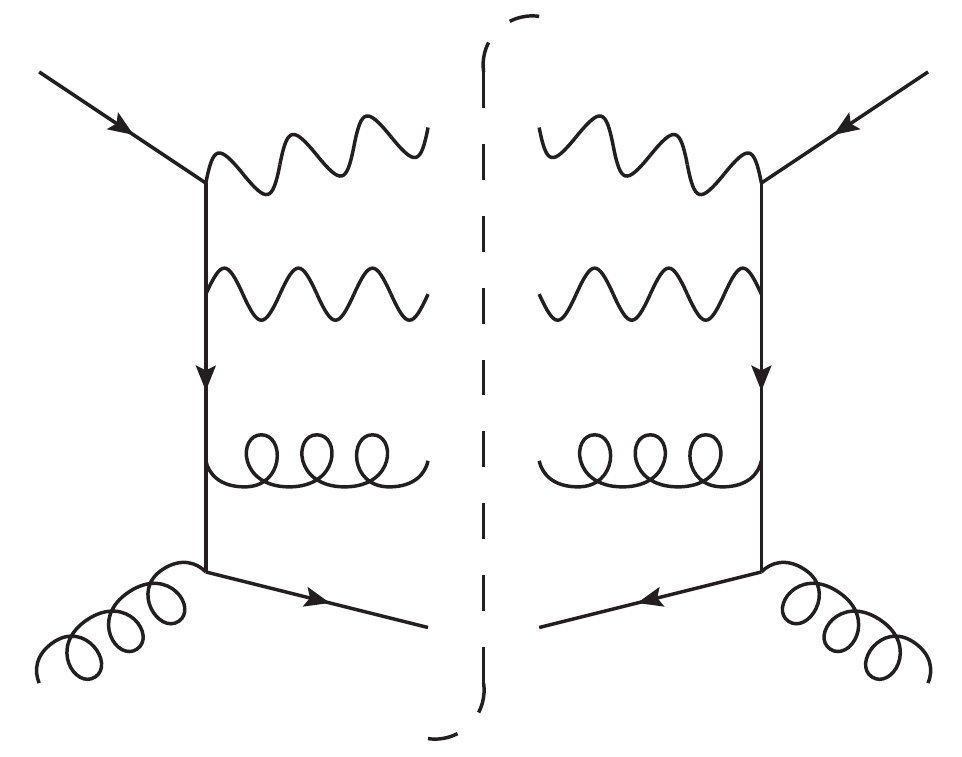} & &
      \includegraphics[width=.20\textwidth]{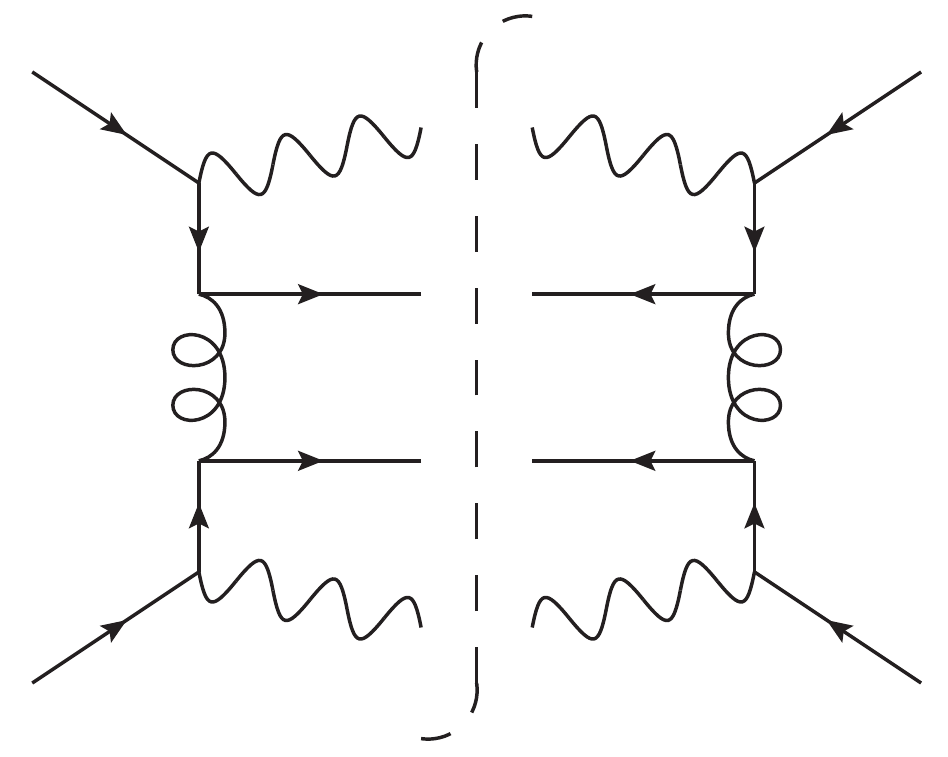} & &
      \includegraphics[width=.20\textwidth]{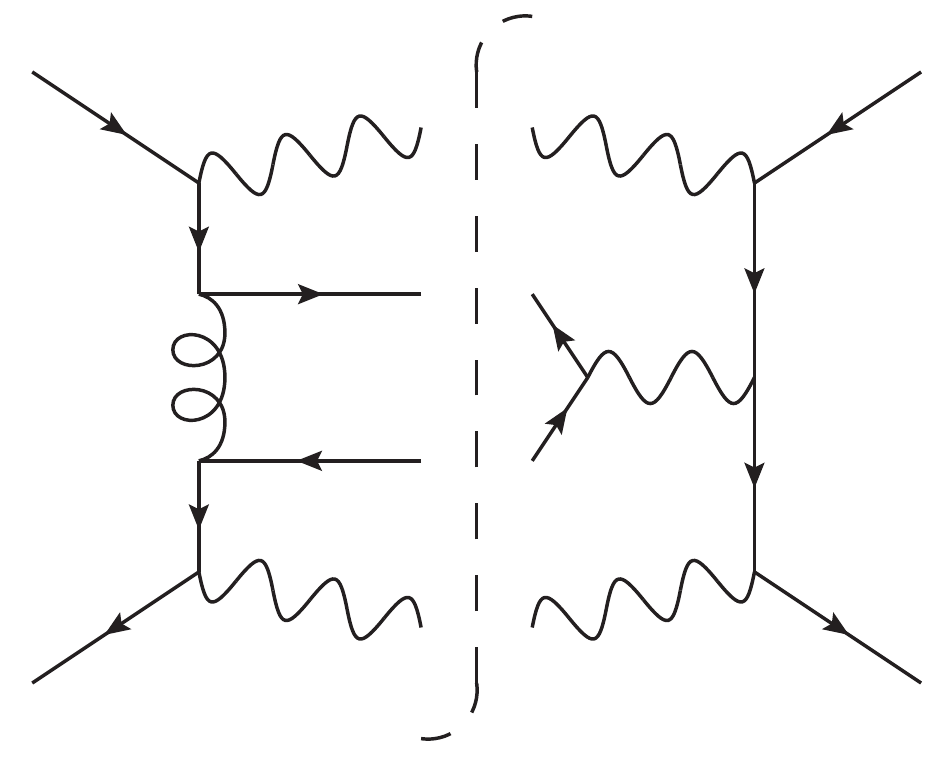} & &
      \includegraphics[width=.20\textwidth]{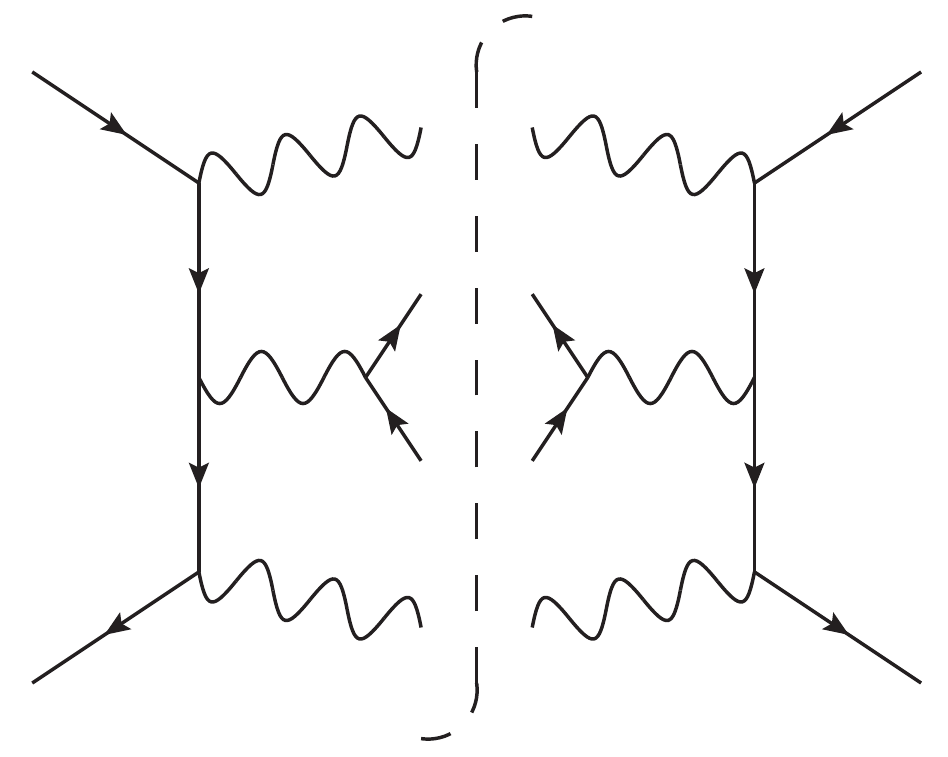} \\[0ex]
      $\order(\alpha_s^2\alpha^2)$ & & 
      $\order(\alpha_s^2\alpha^2)$ & & 
      $\order(\alpha_s\alpha^3)$ & & 
      $\order(\alpha^4)$ 
    \end{tabular}
  \end{center}
  \caption{
    Illustrative leading and sub-leading tree-level diagrams for $\aapa+jj$ production.
    \label{fig:diagrams-aajj-tree}
  }
  \begin{center}
    \small
    \begin{tabular}{ccccccc}
      \includegraphics[width=.20\textwidth]{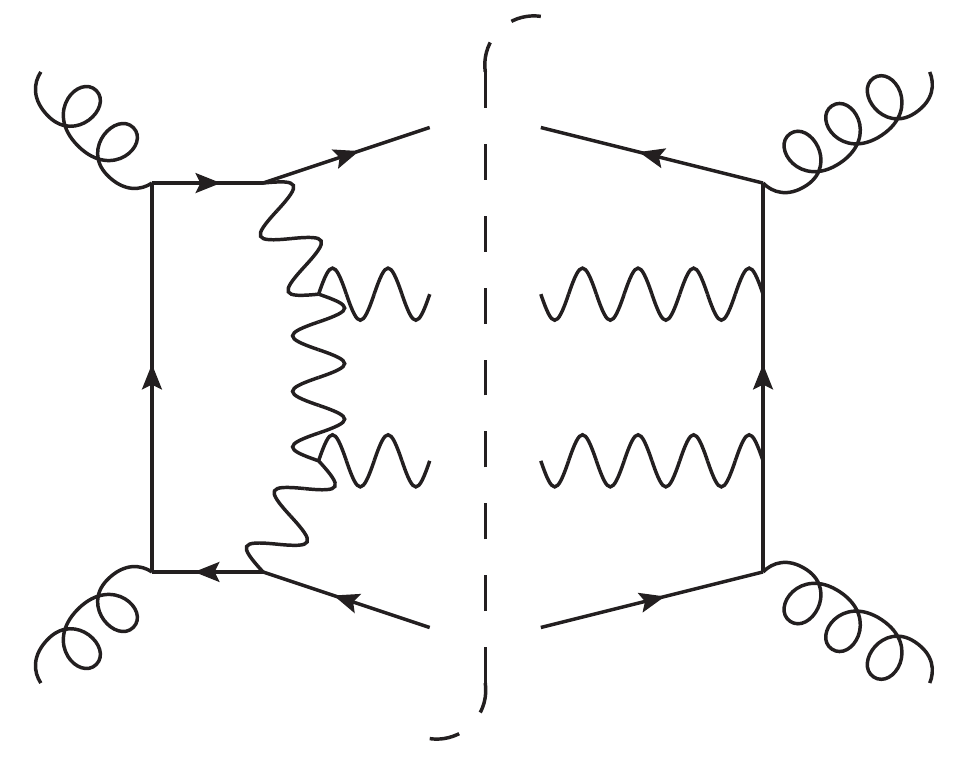} & &
      \includegraphics[width=.20\textwidth]{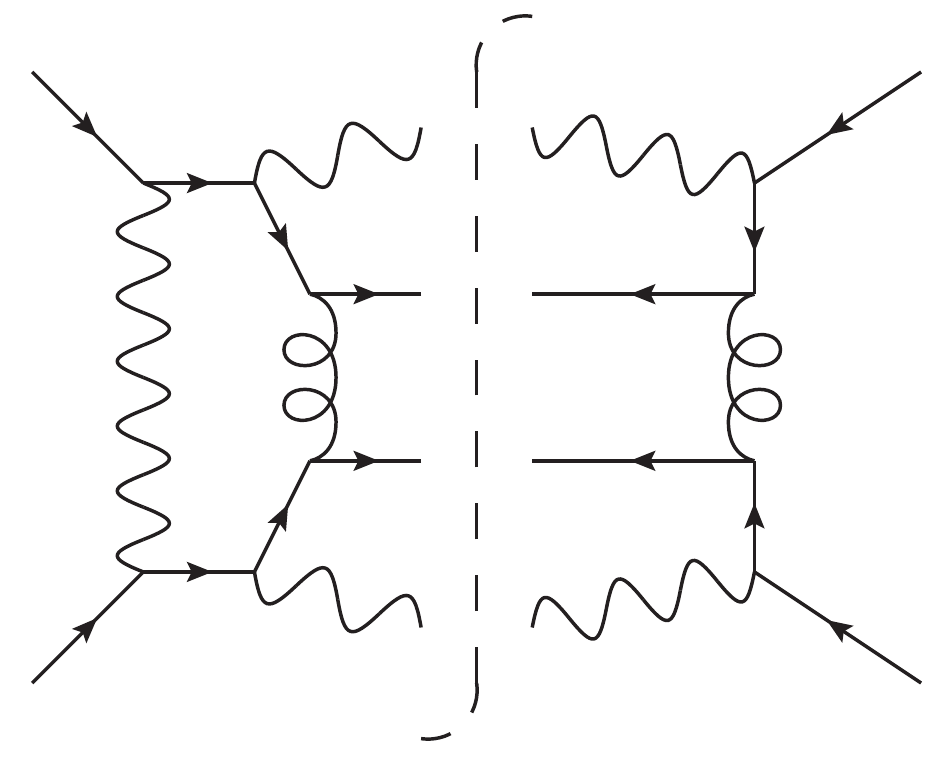} & &
      \includegraphics[width=.20\textwidth]{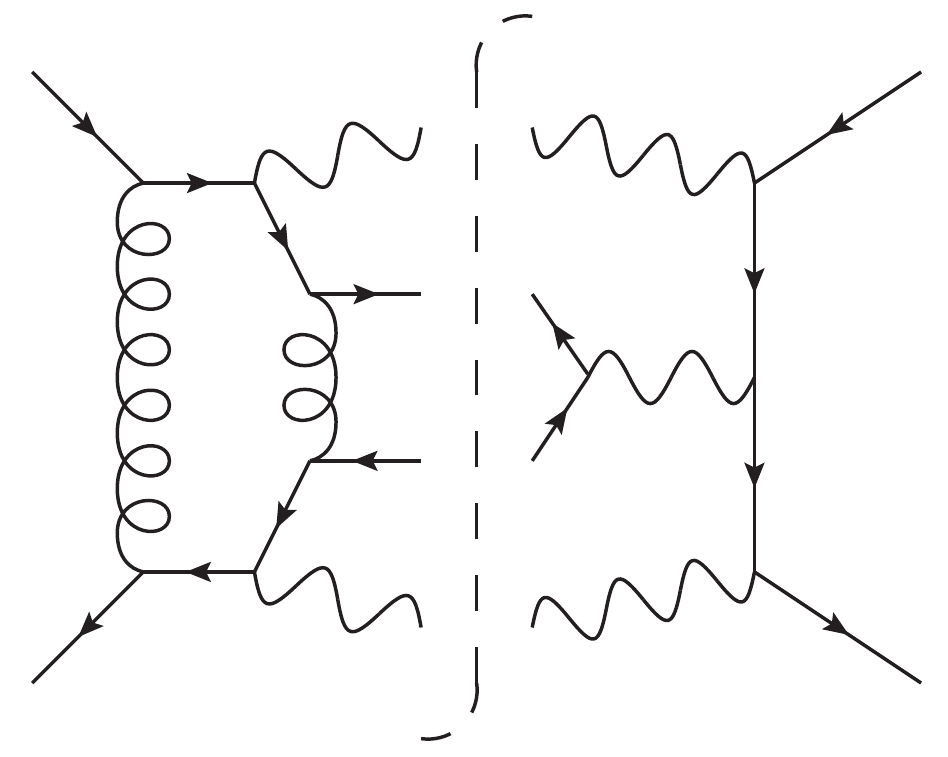} & &
      \includegraphics[width=.20\textwidth]{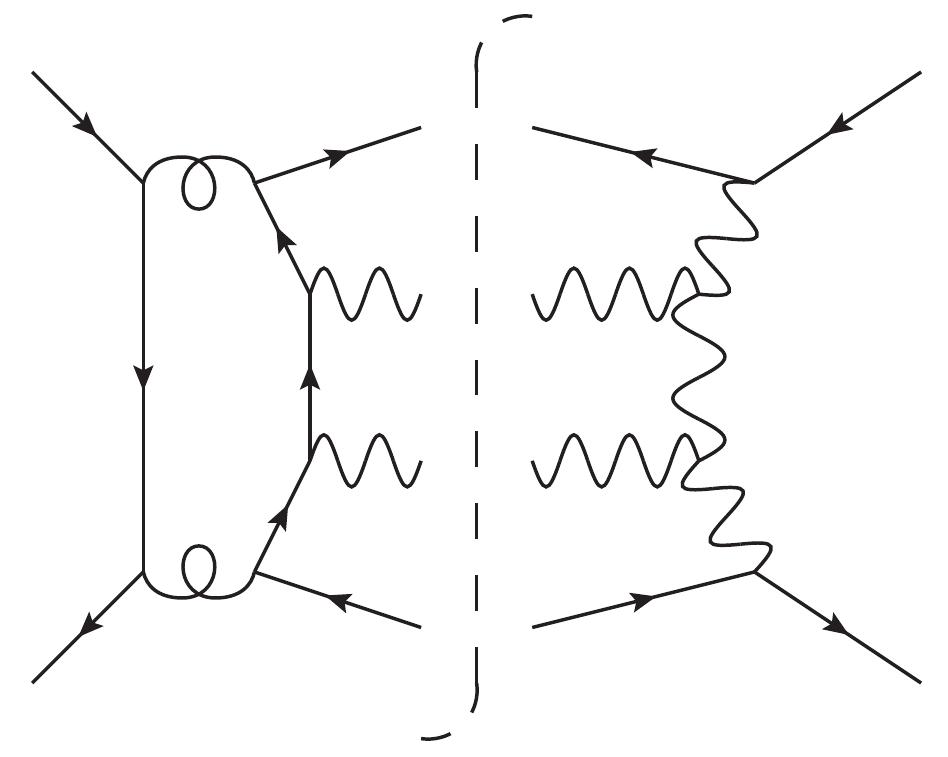} \\[0ex]
      $\order(g_s^2e^4)\times\order(g_s^2e^2)$ & &
      $\order(g_s^2e^4)\times\order(g_s^2e^2)$ & &
      $\order(g_s^4e^2)\times\order(e^4)$ & &
      $\order(g_s^4e^2)\times\order(e^4)$
    \end{tabular}
  \end{center}
  \caption{
    Representative virtual corrections at $\order(\alpha_s^2\alpha^3)$ 
    for $\aapa+jj$ production.
    \label{fig:diagrams-aajj-loop}
  }
\end{figure}

The subleading Born contributions at $\order(\alpha_s\alpha^3)$ 
arise from interferences of ``QCD'' production diagrams at 
$\order(g_s^2e^2)$ and ``EW'' production diagrams at $\order(e^4)$, 
as detailed in Fig.\ \ref{fig:diagrams-aajj-tree}. 
As such, they are not a proiri positive definite. 
Similar considerations arise for the virtual corrections at 
$\order(\alpha_s^2\alpha^3)$, which include pieces that are both 
either insertions of electroweak loops into tree-level processes 
of $\order(\alpha_s^2\alpha^2)$ or insertions of QCD loops into 
interference contributions of $\order(\alpha_s\alpha^3)$, cf.\ 
Fig.\ \ref{fig:diagrams-aajj-loop}.

The contributions at $\order(\alpha^4)$ are pure absolute squares 
of diagrams at $\order(e^4)$ and we will denote the ``sub-subleading'' 
in the following. 
Their contribution is to be considered with care as a significant, 
if not dominant, fraction of the contribution arises from near 
resonant $\gamma\gamma V^*$ ($V=W,Z$) production with subsequent hadronic 
decays of the vector boson. 
At this point, it is a matter of definition of the analysis strategy 
whether these processes are to be included as signal or whether 
they are treated as background. In the latter case they, of course, 
would not be considered to be contributing to the 
$\gamma\gamma+2\,$jets signal. 
Similarly, they could be effectively removed by vetoing events 
with dijet invariant masses in a generous window around the 
nominal electroweak vector boson masses.

\begin{figure}[t!]
 \centering
  \includegraphics[height=0.32\textheight]{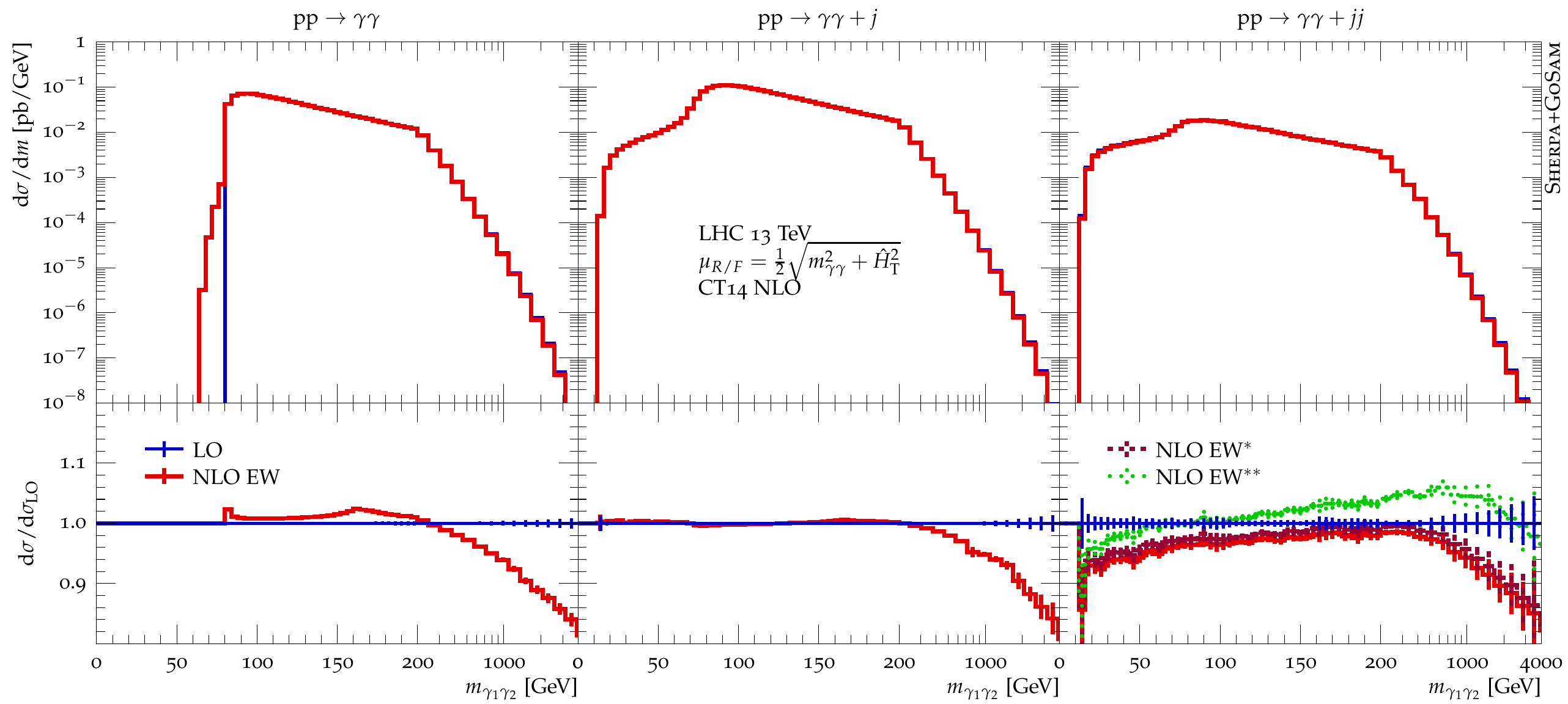}
  \caption{\label{fig:m_yy}
    The invariant mass of the leading and subleading isolated photons 
    in $\aapa+0,1,2$\,jets production at 13\,TeV at the LHC.
  }
\end{figure}

We start our discussion of the differential distributions with the invariant mass distribution of the two leading photons.
 This is shown in Fig.~\ref{fig:m_yy}.
The $0$-jet process is a somewhat special case as at LO the invariant mass is strictly determined by the transverse 
momentum of the two photons and the back-to-back kinematic of the leading order configuration. This back-to-back requirement
is then relaxed by the real emission, but it requires an additional partonic jet at leading order to allow for small invariant masses
of the diphoton system. 
For the $0$-jet process one obtains a small positive correction for low invariant masses with a small peak at $M_W$
 and at $2\cdot M_W$.
It is these peaks that lead to the overall positive correction to the total cross section. Looking more closely at the ingredients
that contribute to the NLO result we see that the peak at $2 M_W$ is generated by the virtual corrections. More precisely
it originates from the class of diagrams
that we illustrated in the left sample diagram of Fig.~\ref{fig:diagrams-aajj-loop} if one just removes the gluons and crosses the 
quarks into the initial state.  The peak at $m_W$ is caused by the real emission contribution and is therefore a threshold
effect, as at leading order the minimal value for the invariant mass is $80$ GeV. For large values of the invariant mass
one obtains the typical decrease the cross section due to the negative Sudakov logarithms of the virtual contributions. For
invariant masses of the order of $1$ TeV one obtains a ten per cent correction. For the $1$-jet process the EW corrections are
essentially zero for small invariant masses and only from $\sim200$ GeV onwards one observes the typical decrease 
which however is irrelevant for the total cross section.  For the $2$-jet process we obtain negative corrections of the order of 
$5-10\%$ already for low invariant masses below the electroweak scale. They reach their minimum for intermediate
scales of up to a few hundred GeV before one finds the decrease of the cross section caused by the Sudakov logarithms.
It is worth mentioning that the subleading Born contribution do not play a significant role, but they are superseded by
the sub-subleading Born contributions. They lead to an overall positive shift which becomes more enhanced for high
invariant masses.  It would be interesting to investigate how their effect would change at NLO EW which is however beyond
the scope of this paper.

\begin{figure}[t!]
  \centering
  \includegraphics[height=0.32\textheight]{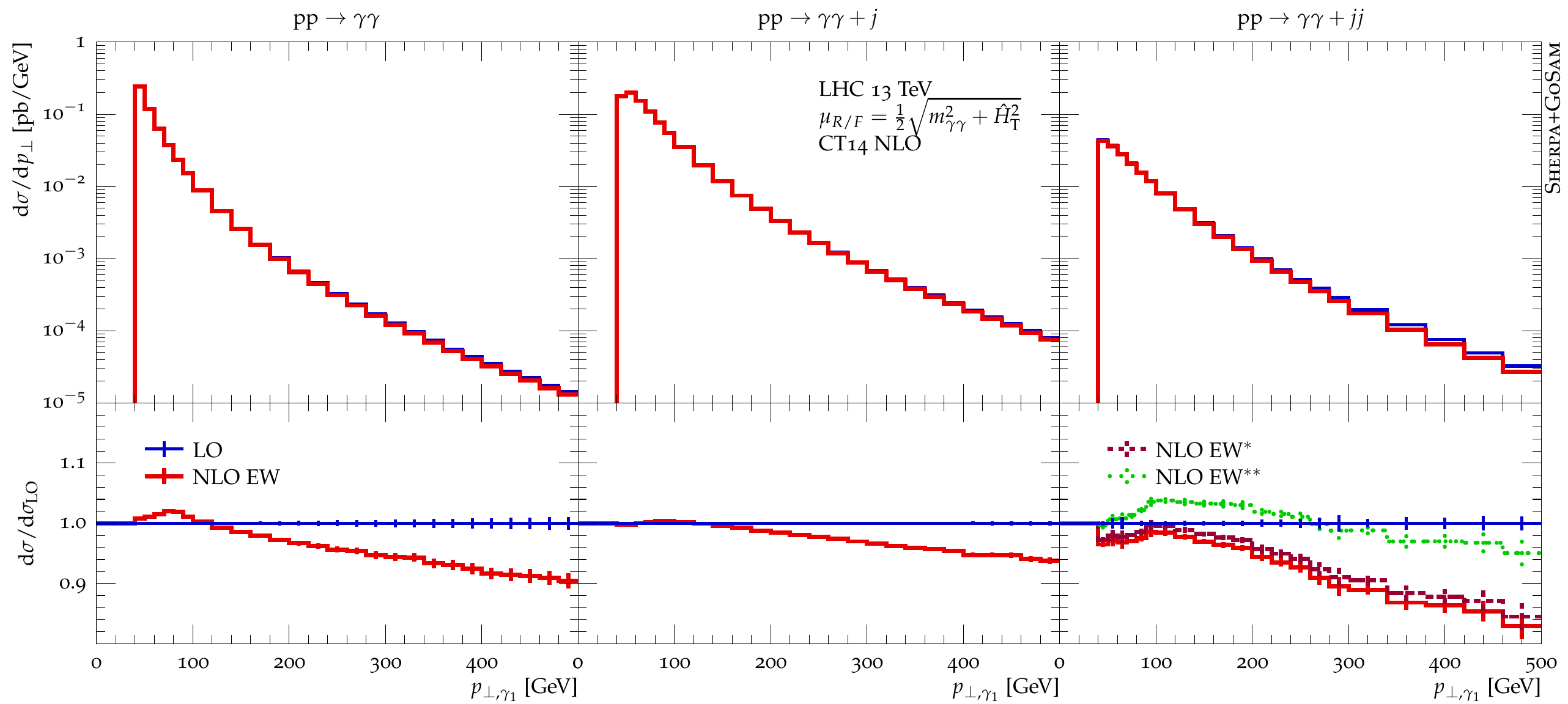}
  \caption{\label{fig:pt_y1}
    The transverse momentum of the leading isolated photon 
    in $\aapa+0,1,2$\,jets production at 13\,TeV at the LHC.
  }
\end{figure}

\begin{figure}[t!]
  \centering
  \includegraphics[height=0.32\textheight]{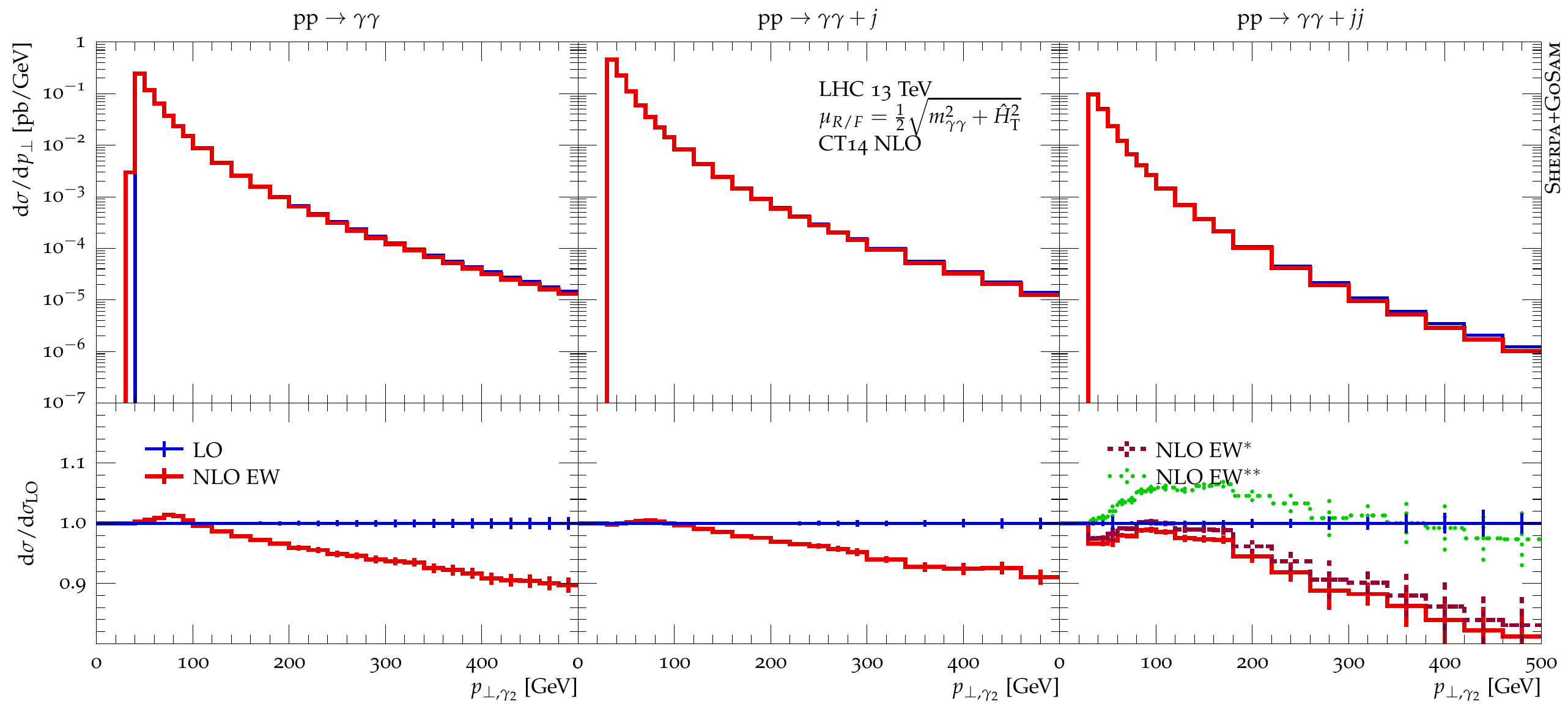}
  \caption{\label{fig:pt_y2}
    The transverse momentum of the subleading isolated photon 
    in $\aapa+0,1,2$\,jets production at 13\,TeV at the LHC.
  }
\end{figure}

A similar pattern can be seen for the transverse momenta of the leading and subleading photon which are shown
in Fig.~\ref{fig:pt_y1} and Fig.~\ref{fig:pt_y2} respectively. The effects of the electroweak corrections affect both photons
in the same way.  For the $0$-jet process one obtains a small positive correction for low transverse momenta up to the 
electroweak scale. The reason for this small positive peak is again the type of virtual diagrams that lead to the peak in
the invariant diphoton mass as discussed above.  The peak is followed by a decrease that leads a $10\%$ correction for 
momenta of $500$ GeV.   For the $1$-jet
process this effect appears to be a bit milder. For the $2$-jet process we observe a negative correction for small
$p_T$ that vanishes around the weak scale and then becomes larger again in the Sudakov regime leading to a $\sim20\%$
correction for large transverse momenta of around $500$ GeV. Also here we see that the subleading born leads to a small
correction and the much bigger effect, in particular for the subleading photon, comes from the sub-subleading Born contribution.
In the tail it leads to an upward shift of almost $20\%$ that essentially nullifies the electroweak corrections to the leading Born.
But as mentioned above this has to be interpreted with care as one then can also expect the electroweak corrections to the
sub-subleading Born to be of non-negligible size.

\begin{figure}[t!]
  \centering
  \includegraphics[height=0.32\textheight]{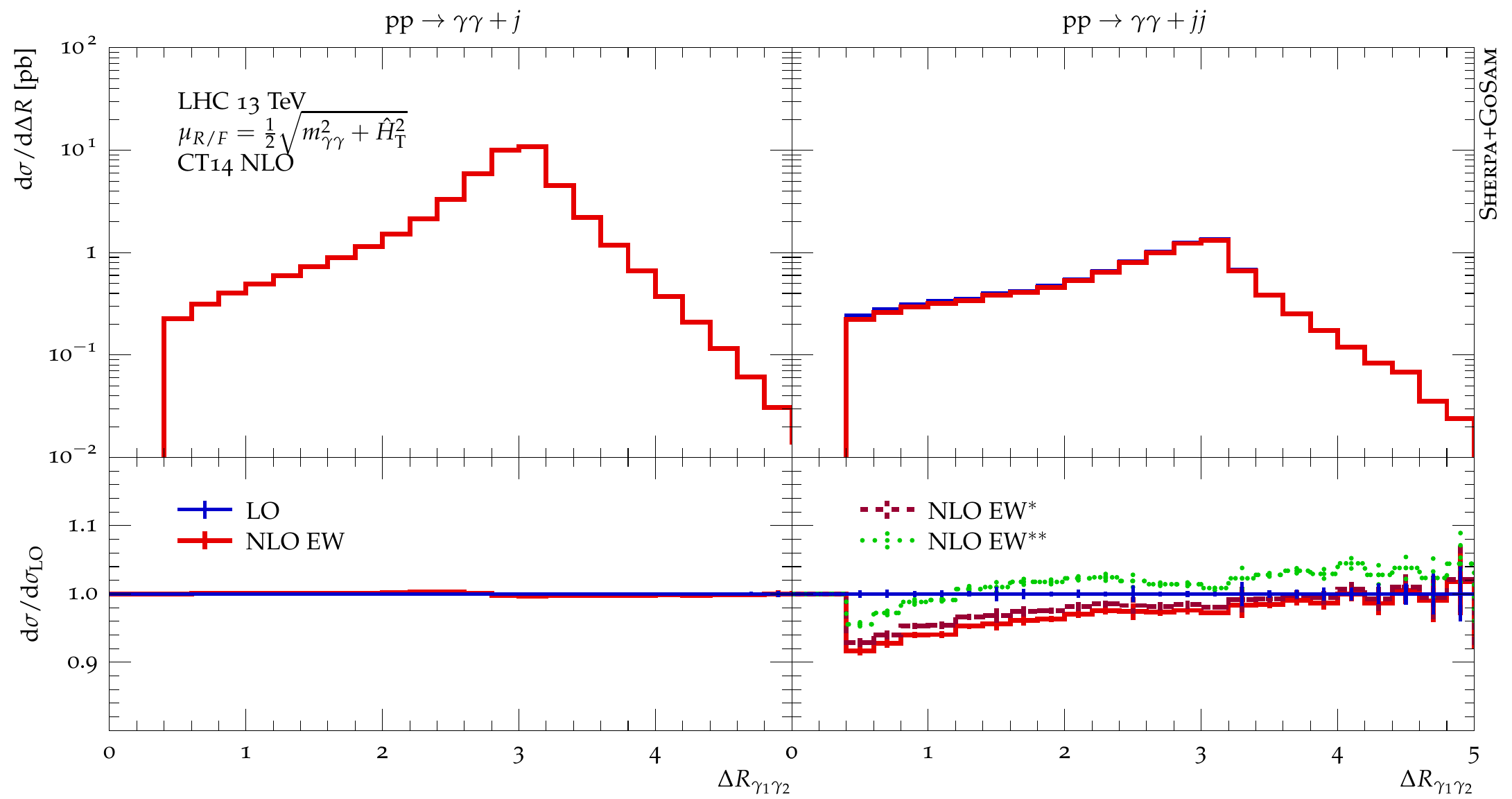}
  \caption{\label{fig:dr}
    Angular separation of the leading and subleading 
    isolated photon system
    in $\aapa+1,2$\,jets production at 13\,TeV at the LHC.
  }
\end{figure}

\begin{figure}[t!]
  \centering
  \includegraphics[height=0.32\textheight]{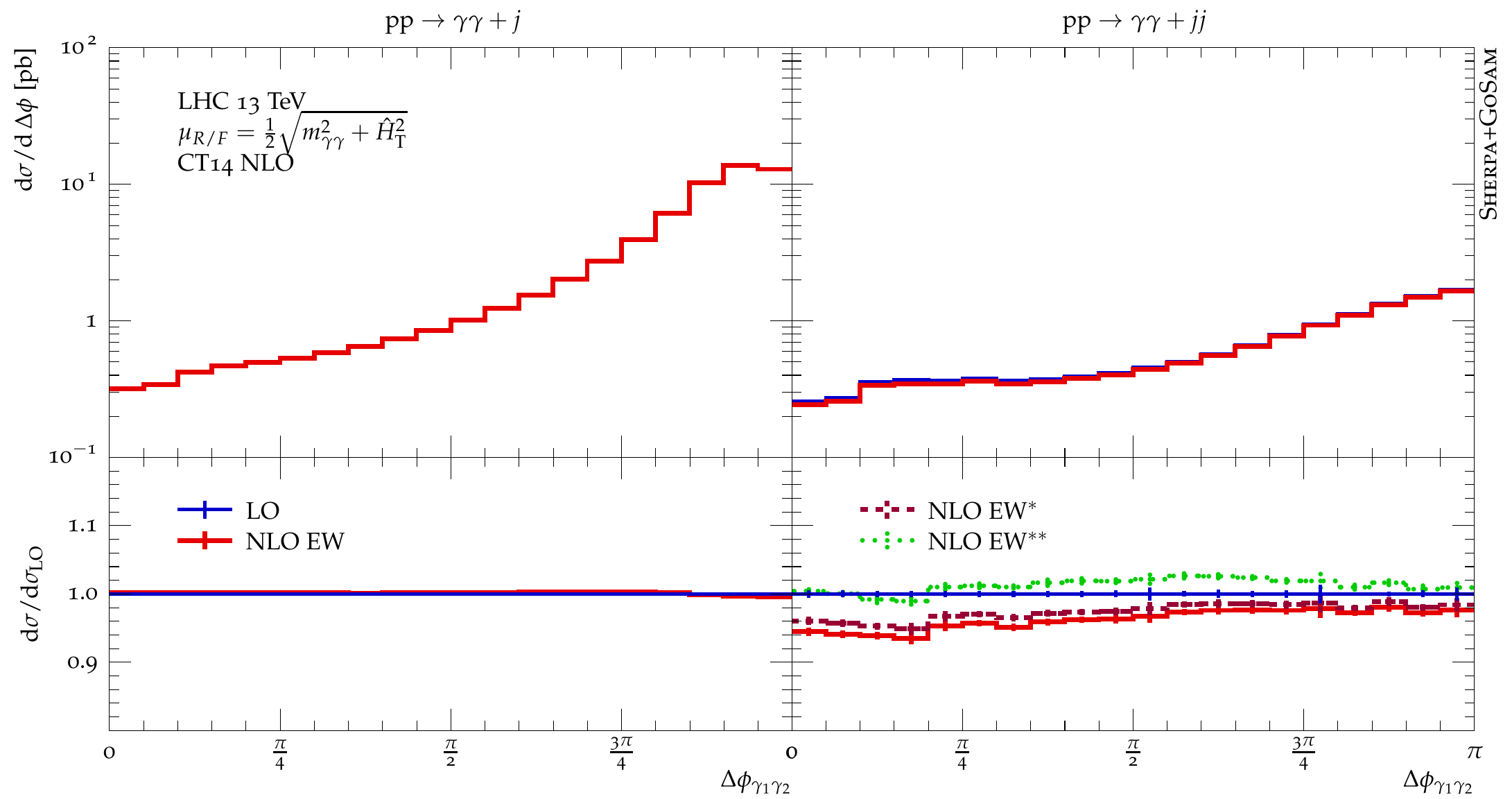}
  \caption{\label{fig:dphi}
    Azimuthal separation of the leading and subleading 
    isolated photon system
    in $\aapa+1,2$\,jets production at 13\,TeV at the LHC.
  }
\end{figure}

The angular separation between the two leading photons is shown in Fig.~\ref{fig:dr}. We do not plot the $0$-jet contribution
as deviations from the back-to-back kinematic are only of leading order accuracy in that case. For the $1$-jet process we find
that the corrections are basically zero. This is not surprising when having in mind that for the total cross section the electroweak
correction were negligible and the corrections are only relevant in the high energy tails. As the angular distributions mix these
regions more or less evenly one cannot expect significant deviations. For the $2$-jet process we find substantial deviations 
of the order of $10\%$ for small values of the angular separation and the corrections become more and more negligible when increasing the angular
separation.

Following this line of arguments it is not surprising that the azimuthal separation which is shown in Fig.~\ref{fig:dphi}, follows
the same pattern as what we have discussed for the angular separation. Also here we find that the corrections in the $1$-jet
process are negligible and for the $2$-jet process we find the largest corrections for small angles and they reach their 
minimum for the maximal separation of $\Delta\phi=\pi$.

\begin{figure}[t!]
  \centering
  \includegraphics[height=0.32\textheight]{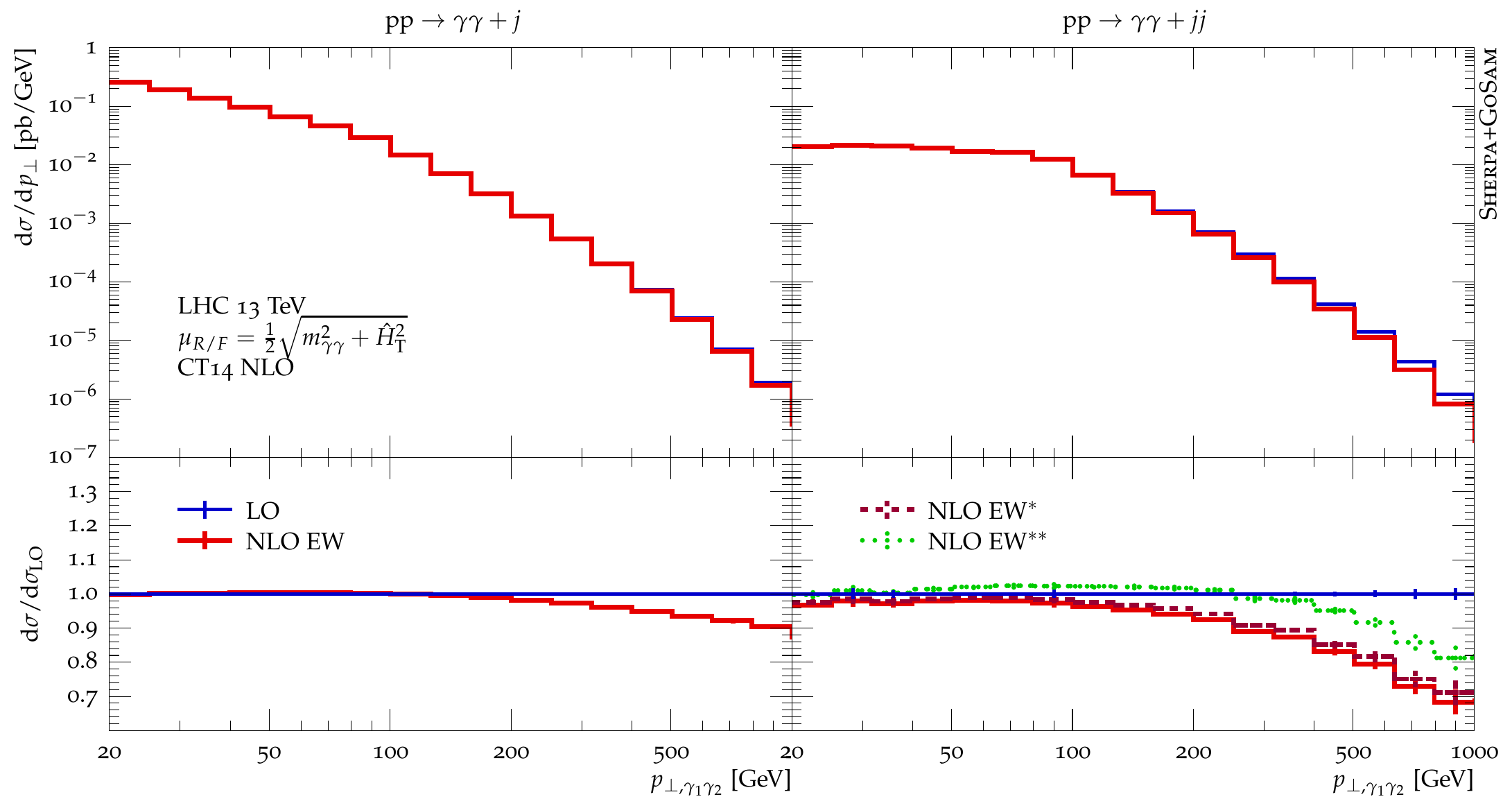}
  \caption{\label{fig:pt_yy}
    The transverse momentum of the leading and subleading 
    isolated photon system
    in $\aapa+1,2$\,jets production at 13\,TeV at the LHC.
  }
  \centering
  \includegraphics[height=0.32\textheight]{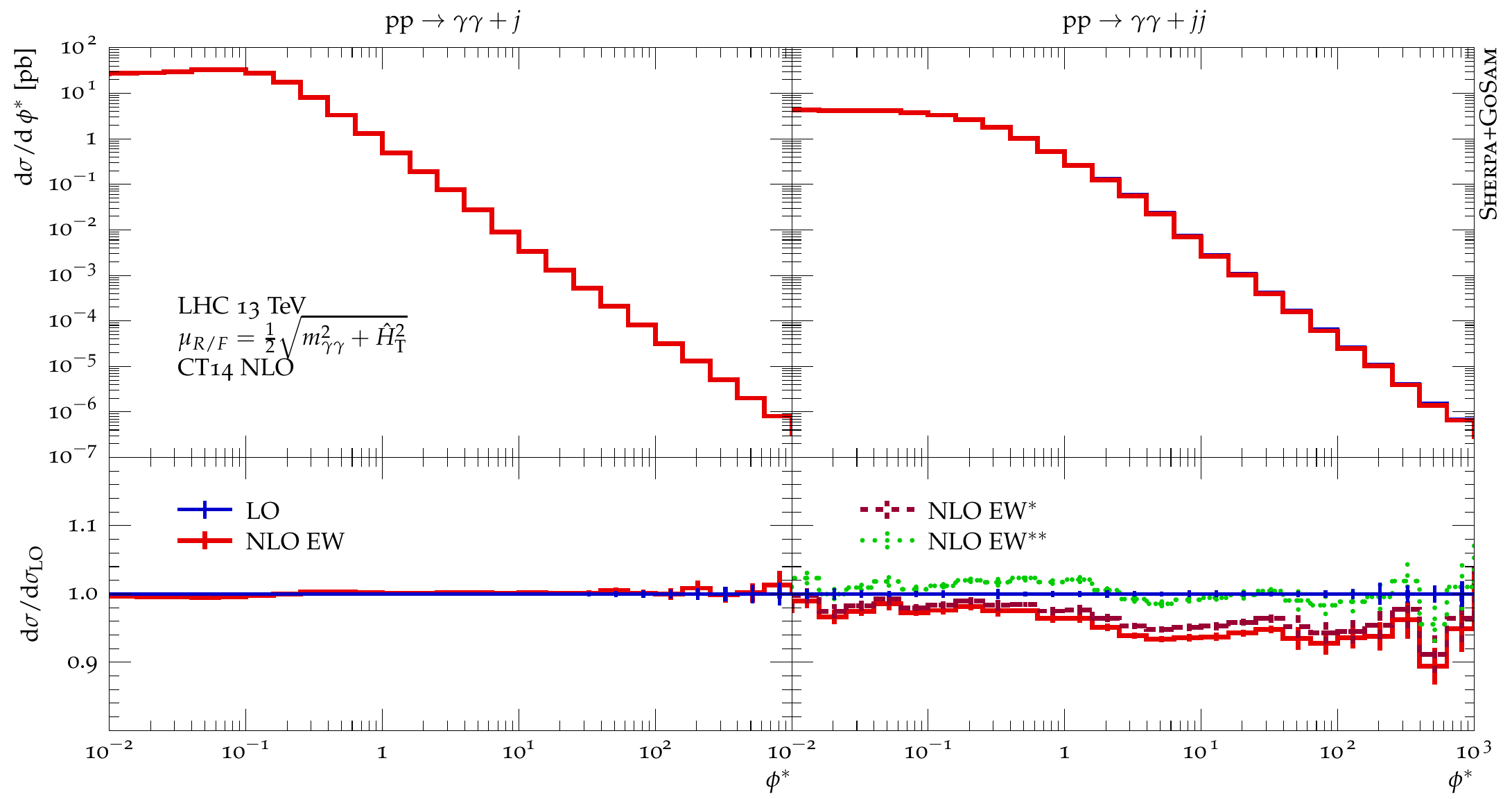}
  \caption{
    $\phi^*$ of the leading and subleading 
    isolated photon system
    in $\aapa+1,2$\,jets production at 13\,TeV at the LHC.
    \label{fig:phistar_yy}
  }
\end{figure}

Another interesting observable, in particular in the context of boosted searches, is the transverse momentum of the diphoton
system. This observable is shown in Fig.~\ref{fig:pt_yy}. Also this observable is only meaningful when having at least one
additional jet so we plot this observable only for the $1$- and $2$-jet process. In both cases the corrections are small for 
small $\pT$ up to the order of $100$ GeV before they become increasingly important leading to a $\sim10\%$ deviation for
the $1$-jet process at high transverse momenta of $1$ TeV. For the $2$-jet process the effects are considerably bigger leading
to a $\sim30\%$ correction. 

\begin{figure}[t!]
  \centering
  \includegraphics[height=0.32\textheight]{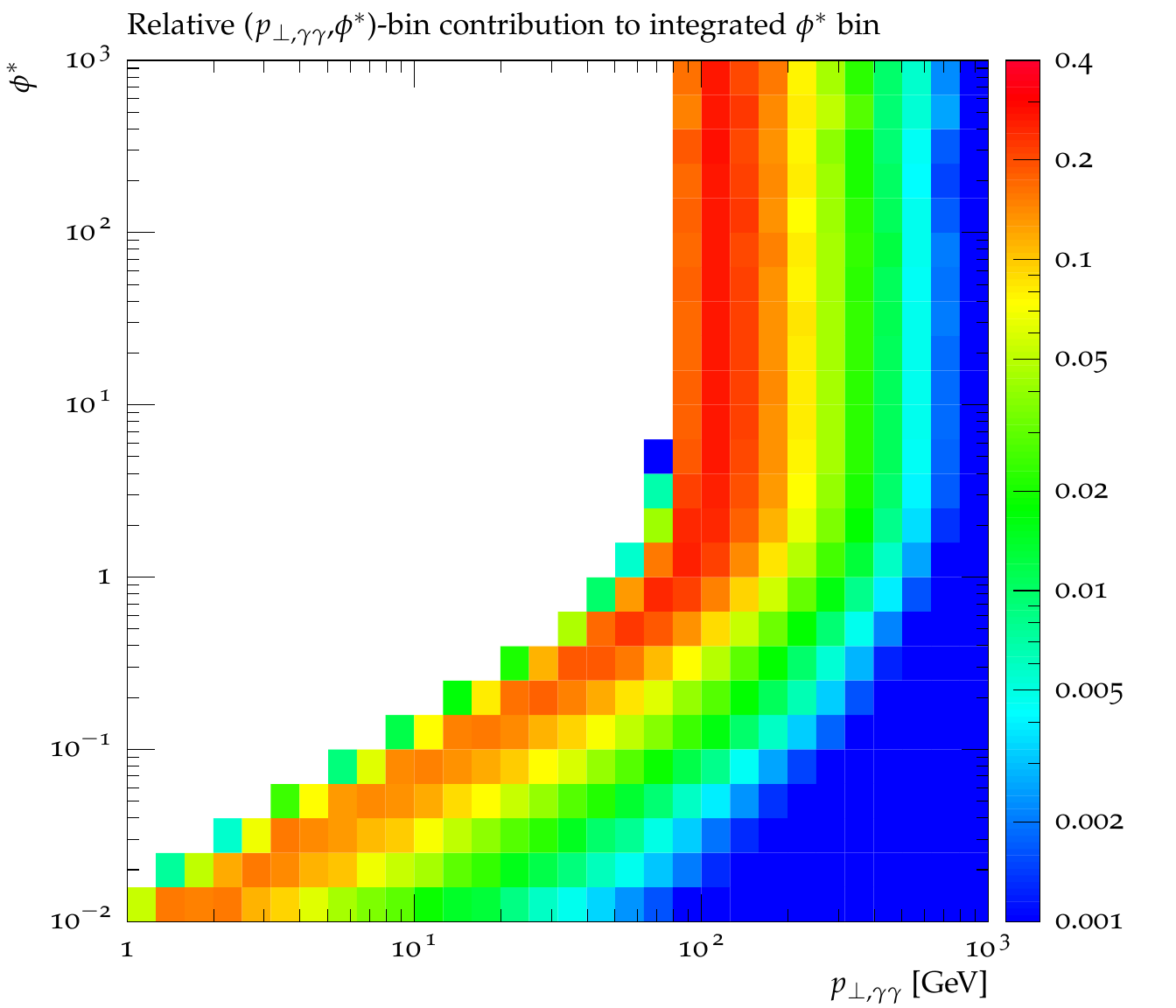}
  \caption{
    Correlation between the transverse momentum of the diphoton system 
    and $\phi^*$ expressed as the fraction each 
    ($p_{\text{T},\gamma\gamma}$,$\phi^*$) bin contributes to the integrated 
    $\phi^*$ bin, calculated at LO accuracy in inclusive $\aapa$ production 
    at 13\,TeV at the LHC. The photon acceptance cuts of eqs.\ 
    \eqref{eq:frix} and \eqref{eq:photon_cuts} are applied, but no jet is 
    required.
    \label{fig:pt-vs-phistar}
  }
\end{figure}

Finally, we investigate the $\phi^*$ observable in Fig.~\ref{fig:phistar_yy}. 
It is defined as \cite{Aaboud:2017vol,Vesterinen:2008hx,Banfi:2010cf}
\begin{equation}
  \phi^* = \tan\frac{\pi-\Delta\phi_{\gamma\gamma}}{2}\cdot\sin\theta^*\;,
\end{equation}
where the diphoton opening angle $\theta^*$ is defined as 
\begin{equation}
  \cos\theta^* = \tanh\frac{\left|\Delta y_{\gamma\gamma}\right|}{2}\;.
\end{equation}
It was introduced to provide an alternative measurement to $\pT$, based 
on well-to-measure particle angles only instead of energy deposits, 
argued to capture the same physics. 
It turns out, however, that the $1$-jet process does not receive any 
correction at all, while the $2$-jet process receives moderate, but 
flat corrections at large $\phi^*$. This behaviour is easy to understand 
since while at least for low transverse momenta $\pT$ and $\phi^*$ are 
very much correlated, this correlation breaks down rather suddenly 
at $\pT\gtrsim 100\,\text{GeV}$ or $\phi^*\gtrsim 1$, 
cf.\ Fig.\ \ref{fig:pt-vs-phistar}.
In this region, any value of $\phi^*$ receives its largest contribution 
from $\pT$ regions around $100$\,GeV. Thus, the electroweak corrections 
can be directly related and vanish for the $1$-jet case. Conversely, 
$\phi^*$ is an ineffective probe of high-$\pT$ physics.

\begin{figure}[t!]
  \centering
  \includegraphics[height=0.32\textheight]{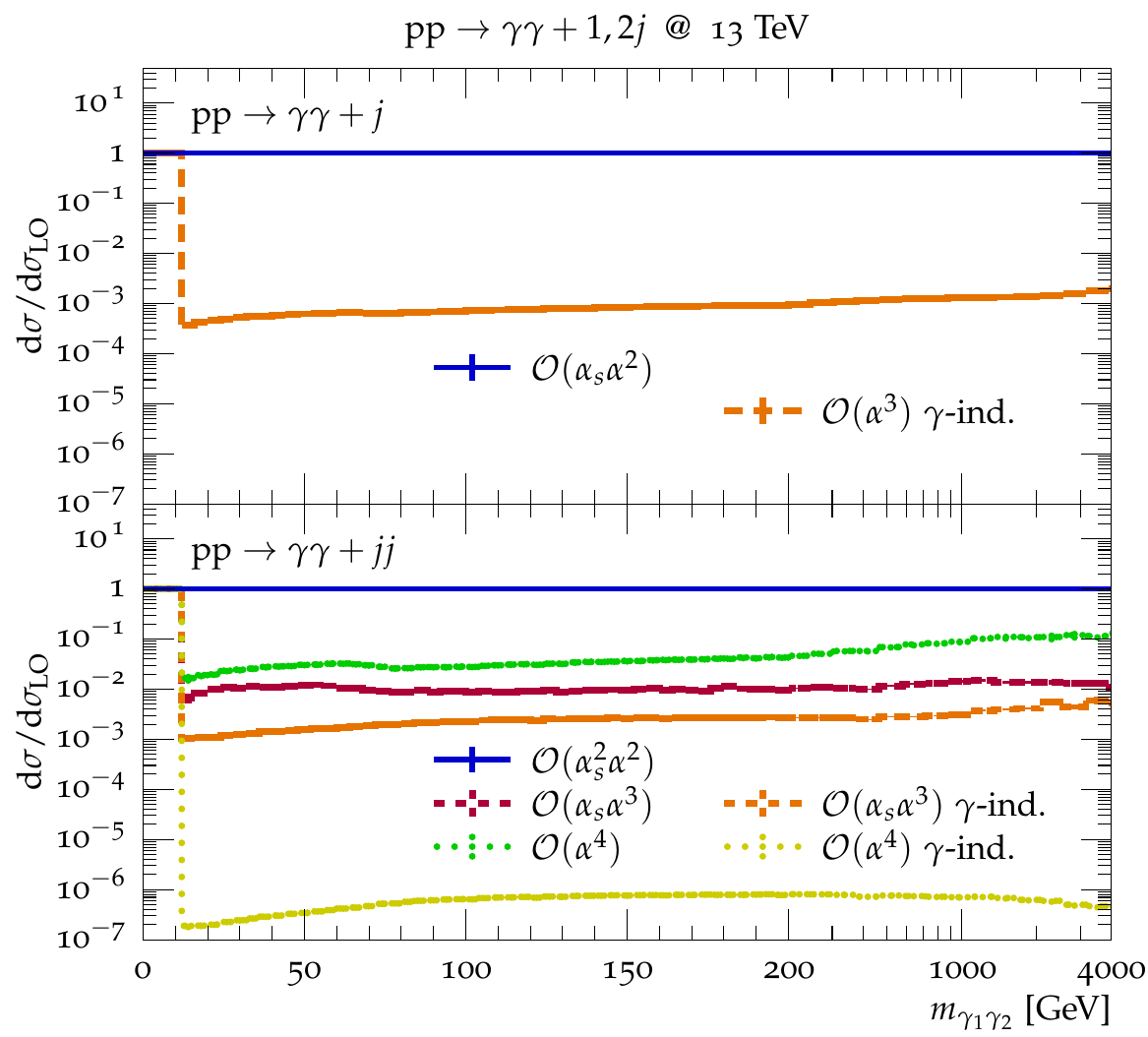}\hfill
  \includegraphics[height=0.32\textheight]{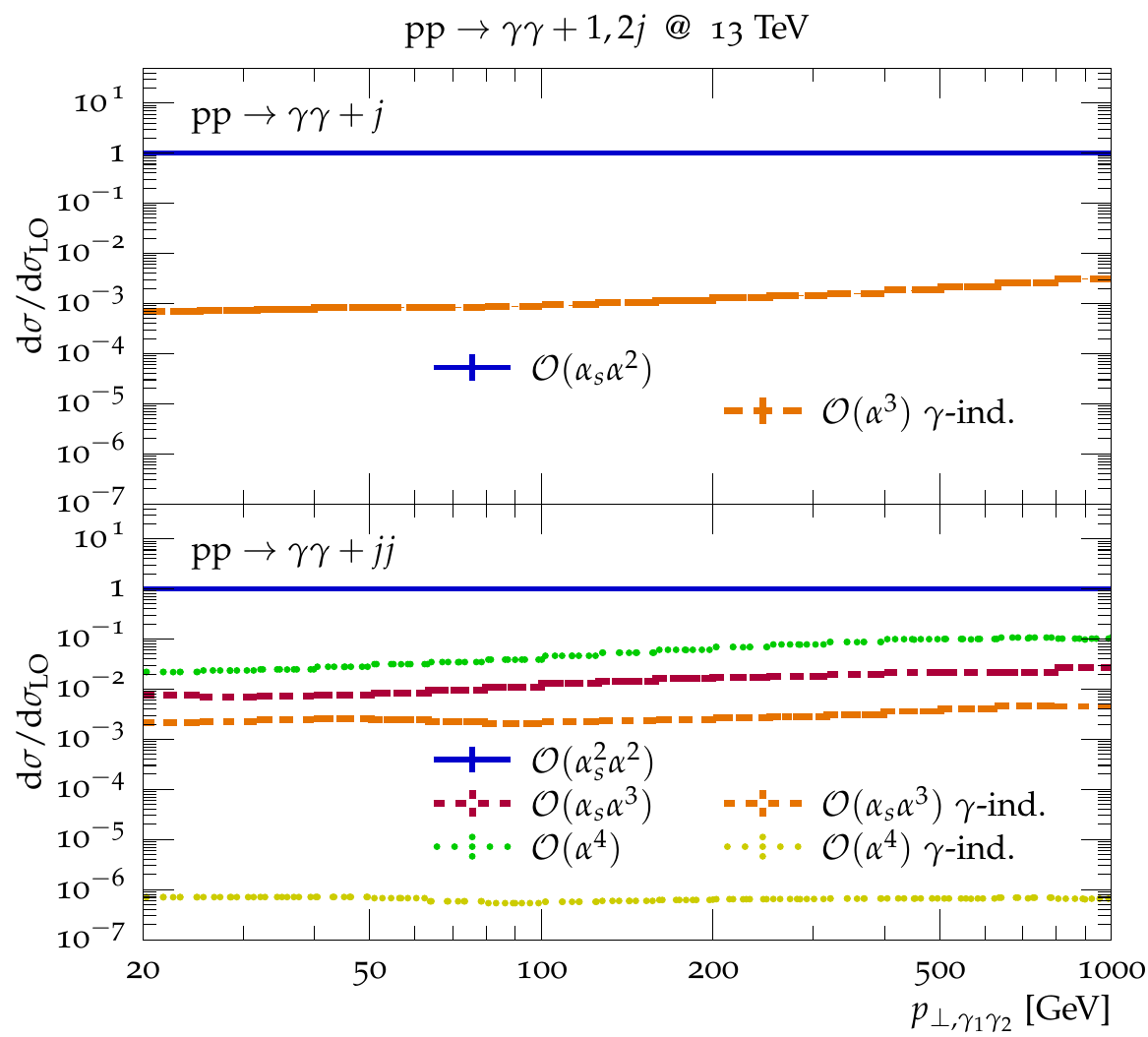}
  \caption{
    Tree-level contributions from subleading orders and photon 
    initiated processes for the invariant mass and the transverse 
    momentum  of the leading and subleading 
    isolated photon system
    in $\aapa+1,2$\,jets production at 13\,TeV at the LHC.
    \label{fig:subleading-contribs}
  }
\end{figure}

Up to this point we have neglected contributions from photon induced 
processes. 
As argued above, they are expected to be small due to the 
smallness of the photon PDF and the fact that these contributions 
only enter at subleading orders at tree-level. 
In the following we investigate the actual size of the contributions 
for two observables where their impact is expected to be largest. 
They are calculated using the CT14nlo QED \cite{Schmidt:2015zda} set 
including both an elastic and an inelastic (with an intrinsic photon 
momentum fraction of $p_0^\gamma=0.005$) component, ensuring the 
quark and gluon PDFs are as close as possible to our nominal set.

Fig.\ \ref{fig:subleading-contribs} details the relative size to the 
leading tree-level contribution of both the subleading and sub-subleading 
and the photon-induced contributions for the diphoton invariant mass 
and the diphoton transverse momentum for both the $\aapa+1\,$jet and 
$\aapa+2\,$jet processes.
For $\aapa+1\,$jet the only subleading contribution is the photon 
induced one, and it contributes well below 0.5\% throughout the 
studied range. 
For $\aapa+2\,$jets the phenomenology is somewhat richer. 
While, as observed above already, the $\order(\alpha^4)$ 
sub-subleading contributions are larger than the 
$\order(\alpha_s\alpha^2)$ ones, they increase the cross section 
at about 10\% and 1\%, respectively. 
The photon-induced processes, on the other hand, again range  
in the sub-percent region, and are completely negligible.